\documentclass[final]{svjour3}                     
\smartqed  
\usepackage{graphicx}
\journalname{Journal of Low Temperature Physics}

\begin{document}

\title{Vortex Shedding from an Object Moving in Superfluid $^4$He at mK Temperatures and in a Bose-Einstein Condensate}


\titlerunning{Vortex Shedding}        

\author{W. Schoepe}


\institute{W. Schoepe \at Fakult\"at f\"ur Physik, Universit\"at Regensburg, Germany \\
              \email{wilfried.schoepe@ur.de}}

\date{}

\maketitle

\begin{abstract}
Vortex shedding from a microsphere oscillating in superfluid $^4$He at mK temperatures is compared with that from a  laser beam moving in a Bose-Einstein condensate (BEC) as observed by other authors. In particular, in either case a linear dependence of the shedding frequency $f_v$ on $\Delta v = v - v_c$ is observed, where $v$ is the velocity amplitude of the sphere or the constant velocity of the laser beam above a critical velocity $v_c$ for the onset of turbulent flow: $f_v = a \,\Delta v$, where the coefficient $a$ is proportional to the oscillation frequency $ \omega $ above some characteristic frequency $\omega_k$ and assumes a finite value for steady motion $\omega \rightarrow 0$.

\end{abstract}

\section*{Introduction}
Quantum turbulence is a common phenomenon in superfluids, ranging from the dense $^4$He and $^3$He liquids to the very dilute Bose-Einstein condensates (BEC). It consists of vortices having a quantized circulation that can be created, e.g., by stirring the superfluids with a moving object or by rotation. In the helium superfluids the easiest way to produce vorticity is by using oscillating objects like spheres, tuning forks or vibrating wires. Because of the simple geometry of a sphere its behavior is more transparent and more easily analyzed than that of the more complicated oscillating structures. 

In the case of a BEC the moving object is typically a laser beam that presents an obstacle to the condensate. The laser beam is swept continuously through the condensate. In addition to the experiments there is a large body of theoretical work on the transition to turbulence based on numerical solutions (mostly 2-dimensional) of the nonlinear Schr\"odinger equation, often known as Gross-Pitaevskii equation, which is applicable for BECs but not for the dense helium liquid.

The motivation for the present article is a comparison of the frequencies at which the vortices are shed in both types of superfluids. We find that in spite of the very different experimental parameters (density, coherence length, speed of sound, interaction strength, linear dimensions, etc.) the shedding frequencies are similar. Moreover, the change from oscillatory flow to steady flow will be discussed for both superfluids.

\section*{The oscillating sphere in superfluid $^4$He}

The behavior of a microsphere (radius $R$ = 0.12 mm) oscillating in superfluid $^4$He at mK temperatures has been investigated at Regensburg University since 1994 \cite {PRL1} and a detailed analysis of the results has been published in a series of reports since then \cite {JLTP135,JLTP173,JLTP178}. Of particular interest is the transition from pure potential flow to turbulence at a critical velocity $v_c$ = 2.8 $\sqrt{\kappa \,\omega }$ (where $\kappa \approx $ 10$^{-7}$ m$^2$/s is the circulation quantum and $\omega$ is the oscillation frequency) \cite{JLTP158}. In a small velocity interval $\Delta v = v - v_c$, where $\Delta v/v_c \le$ 3 $\%$, an instability of the flow pattern is observed, switching intermittently between phases of potential flow and turbulent phases
\cite{JLTP135,JLTP173,JLTP178}, see Fig.1.
\\
\begin{figure}[ht]
\centerline{\includegraphics[width=0.8\textwidth]{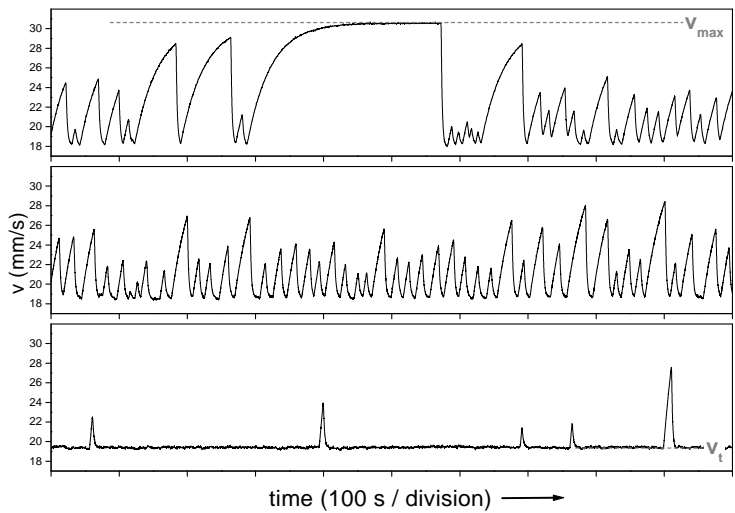}}
\caption{(From Ref.2) Three time series of the velocity amplitude at 300 mK and 114 Hz at three different driving forces (in pN): 47, 55, 75 (from top to bottom). The low level $v_t$ corresponds to turbulent drag while the increase of the velocity amplitude occurs during potential flow, occasionally reaching the maximum value $v_{max}$ given by ballistic phonon drag. Note the rapid increase of the lifetimes of the turbulent phases with the driving force.} 
\label{fig:1}       
\end{figure}

By analyzing the lifetimes of the turbulent phases we find that they are exponentially distributed as exp(-$t/\tau$) and the mean lifetimes $\tau$ grow very rapidly with increasing driving force $F$:
\begin{equation}
\tau = \tau_0 \, \exp[\,(F/F_1)^2\,],
\end{equation}
where $\tau_0$ = 0.5 s for $\omega/2\pi $ = $f $= 119 Hz and 0.25 s for 160 Hz. From a fit of (1) to the data we obtain the force $F_1 = 1.3\, \rho \, \kappa \,R \,\sqrt{\kappa \,\omega}$,
where $\rho$ is the density of liquid $^4$He. This expression is identified as the drag force resulting from the shedding of one vortex ring of size $R$ per half-period (for details, see \cite{JLTP173}). Hence, we conclude that the number $n \equiv F/F_1$ can be interpreted as the average number of vortex rings that are shed during one-half period. We write
\begin{equation}  
\tau^* \equiv \tau/ \tau _0 = \,\exp\, (n^2).
\end{equation}
In our experiments $n$ lies in the interval 0.7 $< n <$ 3, see Fig.2.

\begin{figure}[h]

\centerline{\includegraphics[width=0.8\textwidth]{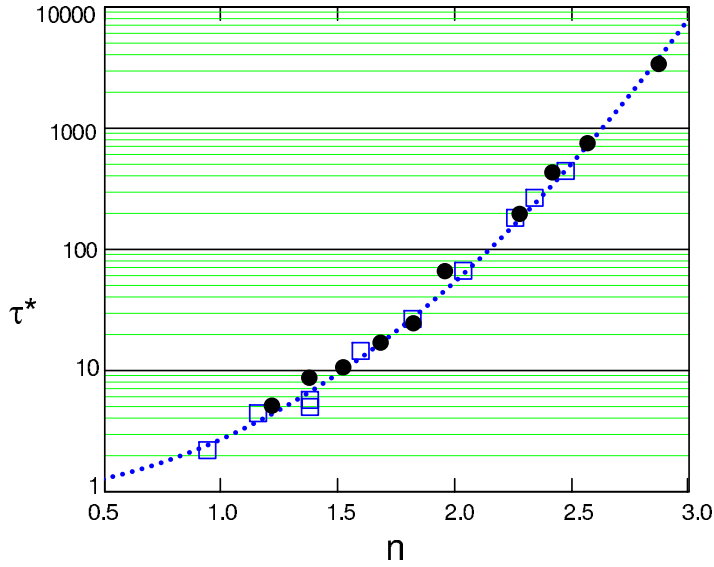}}
\caption{(Color online) The normalized lifetimes $\tau^*$ as function of the average number $n$ of vortex rings shed during one-half period. Blue squares: oscillation frequency 119 Hz, temperature 301 mK. Black dots: 160 Hz, 30 mK, with 0.05$\%\,\,^3$He. Note that the data are independent of oscillation frequency, of temperature, and of $^3$He impurities. The dotted line is calculated from Eq.2.}
\label{fig:2} 
\end{figure}

Replacing the driving force $F$ by the measured drag force \cite{PRL1,JLTP135,JLTP173,JLTP178} $F_d \propto (v^2 - v_c^2) \approx 2 v_c \Delta v$ we find $n = \Delta v /v_1$, with a ``characteristic" velocity $v_1$ = 0.48 $\kappa/R$ = 0.40 mm/s, which is of the same order as the velocity of a vortex ring of radius $R$ \cite{JLTP173}.\\
From $n$ we obtain the average shedding frequency
\begin{equation}
f_v = 2\,n\,f = \frac{2f}{v_1}\Delta v = a\, \Delta v,
\end{equation}
where the coefficient $a$ = 2$f /v_1$. At $f$ = 119 Hz we obtain $a$ = 0.60 $\mu$m$^{-1}$ and at 160 Hz $a$ = 0.80 $\mu$m$^{-1}$. $1/a$ is a relevant length scale that is given here by $v_1 / 2f$, and which can be interpreted as the distance a vortex ring travels during one-half period.

The linear scaling of $a$ with the oscillation frequency is likely to break down at very large frequencies, i.e., when the period becomes shorter than the time it takes a vortex ring to be shed. Then our picture of shedding of individual rings will no longer be applicable. Reconnections and annihilations near the surface of the sphere will become important. In the opposite limit, when the frequency goes to zero (steady motion), it is clear that the shedding frequency must remain finite because vortices can clearly be created also in the case of steady motion. Therefore, there must be a crossover from the linear frequency dependence to a constant value. In that case, the relevant length scale $v_1 / 2f$ is assumed to become equal to the radius $R$ of the sphere. This will happen at a frequency $f = v_1/2R = 0.24\, \kappa / R^2$ = 1.7 Hz which is almost 2 orders of magnitude smaller than our oscillation frequencies. From $a = 1/R$ we find $a = 0.008 \mu$m$^{-1}$, a small but finite value instead of zero. For a sketch of the frequency dependence of the coefficient $a$, see Fig.3.

\begin{figure}[h]

\centerline{\includegraphics[width=0.7\textwidth]{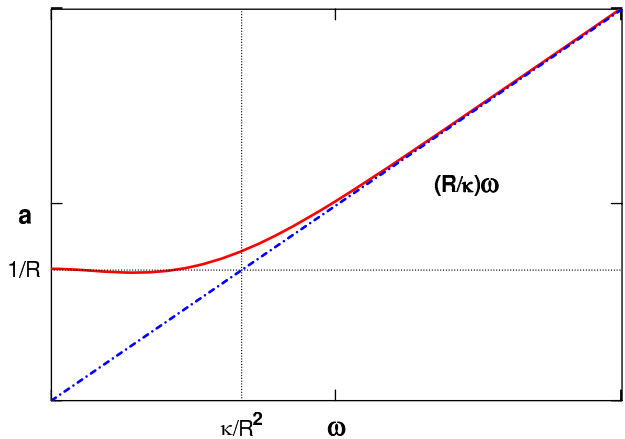}}
\caption{(Color online) Sketch of the coefficient $a$ of $f_v = a \Delta v $ as a function of the oscillation frequency $\omega$. At small frequencies the radius $R$ of the sphere is the characteristic length scale, hence $a \sim 1/R$, whereas at large frequencies $a$ scales as $(R/\kappa)\,\omega$, see (3). The characteristic frequency that marks the transition between both regimes is given by $\kappa /R^2$. (Numerical factors of order 1 are neglected.)}
\label{fig:3} 
\end{figure}

These considerations are similar to those we used to estimate the change of the critical velocity for oscillatory flow $v_c \sim \sqrt{\kappa \omega}$ when the oscillation frequency approaches zero for steady motion. In that case the characteristic length scale changes from the oscillation amplitude to the size of the sphere \cite{JLTP158}, reproducing the Feynman critical velocity $\sim \kappa /R$. The transition occurs at a characteristic oscillation frequency $\omega_k$ when $\sqrt{\kappa \omega_k} \sim \kappa /R$, i.e., at $\omega_k \sim \kappa/ R^2$. It is interesting to note, that $\omega_k$ is the same (except, may be, for some numerical factor) as the frequency where the shedding $f_v$ has its transition, see above. Thus, our experiments with the sphere in $^4$He are clearly in the regime of oscillatory flow for both the critical velocity $v_c$ and the shedding frequency $f_v$.

\section*{Moving a laser beam through a BEC} 
Quantum turbulence has been observed recently at Seoul National University by moving a repulsive Gaussian laser beam 
steadily through a BEC of $^{23}$Na atoms \cite{Shin}. The authors observed shedding of vortex dipoles in a stable and periodic manner. The shedding frequency $f_v$ is found to increase linearly with $\Delta v $: $f_v = a \Delta v$, where $v_c$ = 0.99 mm/s and $a$ = 0.25 $\mu$m$^{-1}$. Because the beam was moved steadily the relevant length scale is the size of the beam 2$R$ = 9.1 $\mu$m, in accordance with the arguments presented above. Therefore we estimate $a \sim 1/R$ = 0.22 $\mu$m$^{-1}$, in fair agreement with the experimental result, and $v_c \sim \kappa /R$ = 3.7 mm/s (where $\kappa$ = 1.7 $10^{-8}$ m$^2$/s for the $^{23}$Na BEC). If the beam would have been oscillating at a frequency substantially larger than $\omega_k \sim \kappa/R^2$ = 803 s$^{-1}$ (or 128 Hz) we would expect a shedding frequency $f_v$ proportional to $\omega$ and a critical velocity scaling as $\omega^{1/2}$, in accordance with our results in $^4$He.

Earlier experiments with a moving beam were performed by Ketterle's group \cite{PRL83,PRL85}. In that work the beam was moved back and forth at frequencies below 200 Hz and the diameter of the beam was 10 $\mu$m. Although the motion was not a sinusoidal one, we can estimate $\omega_k  \sim $ 680 s$^{-1}$ (or 108 Hz), which is of the same order of magnitude as the applied frequencies and, therefore, the frequencies are too close to the steady regime to show the frequency dependencies of both $f_v$ and $v_c$. More details of the work in \cite{PRL83,PRL85} can be found in \cite{JLTP122} where care was taken to distinguish between the critical velocity for heating the BEC due to phonon emission by the moving beam and that of vortex production. In our experiments in helium the situation is simpler because of the very different ratio of the critical velocity to the speed of sound that is of the order of $10^{-4}$ in $^4$He while in the BECs it is typically 0.1 to 0.5. Therefore, in our case the critical velocity is not affected by phonon emission.

Numerical work \cite{Jackson} confirmed the linear increase of the energy dissipation above $v_c$ which is equivalent to a linear increase of $f_v$. That result agrees also with earlier numerical work on the drag force at steady motion \cite{Frisch} from which also a linear behavior of $f_v$ can be inferred.

More recently, an oscillating object in a BEC has been investigated theoretically \cite{Makoto}, but shedding frequencies have not been calculated so far. The frequency range in that work was limited to 0.2$<\omega/\omega_k<$1.0 and, therefore, is also in the regime of steady flow, which is also evident from the frequency independence of the critical velocity for multiple vortex pair production $\approx \kappa /R $.

A very recent work by V.P. Singh et al. \cite{Singh} presents exciting details of the effect of laser stirring of a BEC.
If the stirrer is repulsive the critical velocity is governed by vortex production while for an attractive one phonon emission is the relevant mechanism for heating. Both circular and linear motion of the beam were used. Vortex shedding frequencies were not an issue in that work but an estimate of the characteristic frequency $\omega_k$ gives a value close to the rotation frequency. Therefore, the flow regime was again near the steady limit.

\section*{Summary}
The similarity of the vortex shedding frequency in both superfluid $^4$He and in a BEC is prominent although the systems are quite different, at least at first sight, one reason is that the circulation quanta and the size of the obstacles both differ by an order of magnitude but the ratio $\kappa /R$ is similar. The major difference is that the oscillation frequency of the sphere is much larger than the characteristic frequency $\omega_k$ whereas the laser beam was moved through the BEC in the steady regime. Another difference is that we find an average shedding frequency of up to 3 vortex rings per half-period in our experiment while in \cite{Shin} a strictly periodic emission is observed. Based on our results with the sphere in $^4$He predictions have been made here for experiments with oscillating laser beams in a BEC at much larger frequencies than the characteristic one. It would be interesting to test these considerations experimentally.

\begin{acknowledgements}
Helpful comments from Risto H\"anninen and Matti Krusius (Aalto University, Finland) are gratefully acknowledged. 
\end{acknowledgements}

\end{document}